\input harvmac
\overfullrule=0pt
\def\Title#1#2{\rightline{#1}\ifx\answ\bigans\nopagenumbers\pageno0\vskip1in
\else\pageno1\vskip.8in\fi \centerline{\titlefont #2}\vskip .5in}

\font\ticp=cmcsc10

%
%
\def\sq2{\sqrt{2}}

\def\s42{ 2^{-{1\over 4} } }

\def\a{\alpha}

\def\cg2{\cos (\pi V)}
\def\sg2{\sin (\pi V)}
\def\cb2{\cos (\delta/2)}
\def\sb2{\sin (\delta/2)}

\font\cmss=cmss10 \font\cmsss=cmss10 at 7pt
\def\IZ{\relax\ifmmode\mathchoice
   {\hbox{\cmss Z\kern-.4em Z}}{\hbox{\cmss Z\kern-.4em Z}}
   {\lower.9pt\hbox{\cmsss Z\kern-.4em Z}}
   {\lower1.2pt\hbox{\cmsss Z\kern-.4em Z}}\else{\cmss Z\kern-.4emZ}\fi}

\def\a{\alpha}

\def\sg{r_0^2 {\rm sinh}^2\gamma }

\def\cg{r_0^2 {\rm cosh}^2\gamma }

\def\[{\left [}
\def\]{\right ]}
\def\({\left (}
\def\){\right )}
%
%
\lref\grla{ R. Gregory and R. Laflamme, Phys. Rev. Lett. {\bf 70} (1993)
2837; Nucl. Phys. {\bf B428} (1994) 399.}
\lref\jpup{J. Polchinski, private communication.}
\lref\rk{S. Ferrara and R. Kallosh, hep-th/9602136.}
\lref\host{G. Horowitz and A. Strominger, Nucl. Phys. {\bf B360} (1991) 197.  }
\lref\hhs{J. Horne, G. Horowitz, and A. Steif, Phys. Rev. Lett. {\bf 68} (1992)
 568, hep-th/9110065.  }
\lref\gkp{S. Gubser, I. Klebanov and A. Peet, hep-th/9602135;
A. Strominger, unpublished.}
\lref\witvar{E. Witten, Nucl. Phys. {\bf B 443} (1995) 85, hep-th/9503124.  }
\lref\cd{M. Cvetic and D. Youm, hep-th/9603100.}
\lref\dbr{J. Polchinski, S. Chaudhuri, and C. Johnson, hep-th/9602052.}
\lref\jp{J. Polchinski, Phys.Rev.Lett.{\bf 75} (1995) 4724,
 hep-th/9510017; 
J. Dai, R. Leigh and J. Polchinski, Mod. Phys.
Lett. {\bf A4} (1989) 2073.}
\lref\dm{S. Das and S. Mathur, hep-th/9601152.}
\lref\witb{E. Witten, hep-th/9510135.}
\lref\vgas{C. Vafa, hep-th/9511088.}
\lref\ghas{G. Horowitz and A. Strominger, hep-th/9602051.}
\lref\bsv{M. Bershadsky, V. Sadov and C. Vafa,
hep-th/9511222.}
\lref\vins{C. Vafa, hep-th/9512078.}
\lref\cvetd{M. Cvetic and D. Youm, hep-th/9507090.}
\lref\chrs{D. Christodolou, Phys. Rev. Lett. {\bf 25}, (1970) 1596;
D. Christodolou and R. Ruffini, Phys. Rev. {\bf D4}, (1971) 3552.}
\lref\cart{B. Carter, Nature {\bf 238} (1972) 71.}
\lref\penr{R. Penrose and R. Floyd, Nature {\bf 229} (1971) 77.}
\lref\hawka{S. Hawking, Phys. Rev. Lett. {\bf 26}, (1971) 1344.}
\lref\sussc{L.~Susskind,  Phys. Rev. Lett. {\bf 71}, (1993) 2367;
L.~Susskind and L.~Thorlacius, Phys. Rev. {\bf D49} (1994) 966;
L.~Susskind, ibid.  6606.}
\lref\polc{J. Dai, R. Leigh and J. Polchinski, Mod. Phys.
Lett. {\bf A4} (1989) 2073.}
\lref\ascv{A. Strominger and C. Vafa, hep-th/9601029.}
\lref\hrva{P. Horava, Phys. Lett. {\bf B231} (1989) 251.}
\lref\cakl{C. Callan and I. Klebanov, hep-th/9511173.}
\lref\prskll{J. Preskill, P. Schwarz, A. Shapere, S. Trivedi and
F. Wilczek, Mod. Phys. Lett. {\bf A6} (1991) 2353. }
\lref\sbg{S. Giddings, Phys. Rev {\bf D49} (1994) 4078.}
\lref\cghs{C. Callan, S. Giddings, J. Harvey, and A. Strominger,
Phys. Rev. {\bf D45} (1992) R1005.}
\lref\cvyo{M. Cvetic and D. Youm, hep-th/9507090.}
\lref\tse{A. Tseytlin, hep-th/9601117.}
\lref\cvpt{M. Cvetic, private communication.}
\lref\bhole{G. Horowitz and A. Strominger,
Nucl. Phys. {\bf B360} (1991) 197.}
\lref\bekb{J. Bekenstein, Phys. Rev {\bf D12} (1975) 3077.}
\lref\hawkb{S. Hawking, Phys. Rev {\bf D13} (1976) 191.}
\lref\wilc{P. Kraus and F. Wilczek, hep-th/9411219, Nucl. Phys.
{\bf B433} (1995) 403. }
\lref\ght{G. Gibbons, G. Horowitz, and P. Townsend, hep-th/9410073,
Class. Quantum Grav.,
{\bf 12} (1995) 297. }
\lref\intrp{G. Gibbons and P. Townsend, Phys. Rev. Lett.
{\bf 71} (1993) 3754.}
\lref\gmrn{G. Gibbons, Nucl. Phys. {\bf B207} (1982) 337;
G. Gibbons and K. Maeda Nucl. Phys. {\bf B298} (1988) 741.}
\lref\bch{J. Bardeen, B. Carter and S. Hawking,
Comm. Math. Phys. {\bf 31} (1973) 161.}
\lref\stas{A.~Strominger and S.~Trivedi,  Phys.~Rev. {\bf D48}
 (1993) 5778.}
\lref\jpas{J.~Polchinski and A.~Strominger,
hep-th/9407008, Phys. Rev. {\bf D50} (1994) 7403.}
\lref\send{A. Sen, hep-th/9510229, hep-th/9511026.}
\lref\cvet{M. Cvetic and A. Tseytlin, hep-th/9512031.}
\lref\kall{R. Kallosh, A. Linde, T. Ortin, A. Peet and 
A. van Proeyen, Phys. Rev. 
{\bf D46} (1992) 5278.}
\lref\lawi{F. Larsen and F. Wilczek, hep-th/9511064; hep-th/9604134.}
\lref\bek{J. Bekenstein, Lett. Nuov. Cimento {\bf 4} (1972) 737,
Phys. Rev. {\bf D7} (1973) 2333, Phys. Rev. {\bf D9} (1974) 3292.}
\lref\hawk{S. Hawking, Nature {\bf 248} (1974) 30, Comm. Math. Phys.
{\bf 43} (1975) 199.}
\lref\cama{C. Callan and J. Maldacena, hep-th/9602043.}
\lref\sen{A. Sen,  Mod. Phys. Lett. {\bf A10} (1995) 2081,
hep-th/9504147.}
\lref\suss{L. Susskind, hep-th/9309145.}
\lref\sug{L. Susskind and J. Uglum, hep-th/9401070, Phys. Rev. {\bf D50}
 (1994) 2700.}
\lref\peet{A. Peet, hep-th/9506200.}
\lref\tei{C. Teitelboim, hep-th/9510180.}
\lref\carl{S. Carlip, gr-qc/9509024. }
\lref\thoo{G. 'tHooft, Nucl. Phys. {\bf B335} (1990) 138
Phys. Scr. {\bf T36} (1991) 247.}
\lref\fks{S. Ferrara, R. Kallosh and A. Strominger, hep-th/9508072,
Phys. Rev. {\bf D 52}, (1995) 5412 .}
\lref\spn{J. Breckenridge, R. Myers, A. Peet and C. Vafa, hep-th/9602065.}
\lref\vbd{J. Breckenridge, D. Lowe, R. Myers, A. Peet, A. Strominger 
and C. Vafa, hep-th/9603078.}
\lref\townsend{ C. Hull and P. Townsend, Nucl. Phys. {\bf B 438} (1995) 109,
hep-th/9410167. }
\lref\scherk{ E. Cremmer, J. Scherk and J. Schwarz, Phys. Lett. {\bf B 84 }
(1979) 83.}
\lref\dgl{ M. Douglas, hep-th/9512077.}
\lref\jmas{ J. Maldacena and A. Strominger, hep-th/9603060.}  
\lref\cjrm{C. Johnson, R. Khuri and R. Myers, hep-th/9603061.}
\lref\hlm{ G. Horowitz, D. Lowe and J. Maldacena, hep-th/9603195. }
\lref\cvetic{ M. Cvetic and D. Youm, hep-th/9508058.}

\lref\hms{G. Horowitz, J. Maldacena and A. Strominger, hep-th/9603109.}

\lref\sm{ J. Maldacena and L. Susskind, hep-th/9604042.}

\lref\verlinde{R. Dijkgraaf,  E. Verlinde and H.  Verlinde, hep-th/9603126;
hep-th/9604055.}

\lref\cveticnew{M. Cvetic and D. Youm, hep-th/9603147.}

\lref\endsoftwobranes{ A. Strominger, hep-th/9512059.}

\lref\thooft{G. 't Hooft, Commun. Math. Phys. {\bf 81 } (1981) 267.}

\lref\klts{ I. Klebanov and A. Tseytlin, hep-th/9604089.}

\lref\hull{
C. Hull and P. Townsend, Nucl. Phys. {\bf B438} (1995) 109,
hep-th/9410167.}

\lref\schwarz{J. H. Schwarz, hep-th/9604171.}

\lref\peet{A. Peet, hep-th/9506200.}

\lref\cmp{C. Callan, J. Maldacena and A. Peet, hep-th/9510134.}

\lref\lsrutgers{ L. Susskind, Rutgers University preprint RU-93-44,
August 1993, hep-th/9309145.}

\lref\dilgra{ J. Harvey and A. Strominger, ``{\it Quantum aspects 
of black holes}'' in Proceedings of the 1992 Theoretical Advanced Study
Institute in Elementary Particle Physics, J. Harvey and J. Polchinski Eds.,
World Scientific.}


%
\Title{\vbox{\baselineskip12pt
\hbox{PUPT-1621} \hbox{hep-th/9605016}}}
{\vbox{
\centerline {Statistical Entropy of Near Extremal
Five-Branes}}}
\vskip.1in
\centerline{{\ticp
 Juan M. Maldacena\foot{
e-mail: malda@puhep1.princeton.edu} }}
\vskip.3in
\centerline{\it Joseph Henry Laboratories, Princeton University,
Princeton, NJ 08544, USA}
\bigskip
\bigskip
\bigskip
\centerline{\bf Abstract}
The Hawking Beckenstein entropy of near extremal 
fivebranes is calculated in terms of a gas of
strings living on the fivebrane. 
These fivebranes can also be  viewed as   near extremal
black holes in five dimensions.
\Date{}
%
Recently
the Bekenstein-Hawking entropy of certain extremal BPS black holes 
was precisely calculated by 
counting microstates in string theory  \refs{\ascv \spn \jmas -\cjrm}.
Since black holes are non-perturbative objects, the calculations
required considerable understanding  of
non-perturbative string theory and a certain class of solitons called
D-branes  \jp.

These results were
 extended to leading order away from extremality in
\refs{\cama \ghas \vbd -\hlm}. 
These calculations are not so well justified as the BPS ones because
one naively expects strong coupling effects in 
the semiclassical regime in which
the black hole picture is valid. Nevertheless it was argued that
strong coupling effects could be avoided in certain corners of the parameter
space sufficiently near to extremality. The surprising
agreement discovered in \refs{\cama \ghas \vbd -\hlm}
 between the string and
black hole calculations indicate that under favorable circumstances
strong coupling effects could be  avoided.
The calculation in this paper suffers a similar  kind
of  strong coupling  problem, but it extends the
ones in \refs{\cama \ghas \vbd -\hlm} to a wider class of black holes, 
finding an interesting interpretation 
of the classical formulas.

In \refs{\gkp,\klts} the entropy of some near extremal black $p$-branes 
was observed to scale as that of  a gas of particles living on the brane. 
The ten dimensional 
 fivebrane that  we consider here  is a different case which did not
fit in the  analysis of  \klts . 
In \verlinde\ the entropy for BPS fivebranes was calculated in 
terms of strings living on the fivebrane, here we try to give  the 
description for non-BPS configurations.

\newsec{Classical solutions }

In ref. \hms\ general non extremal black hole solutions to
 type II supergravity
compactified on $T^5 = T^4\times S_1$ 
were considered. These black holes are characterized
by three
independent charges, their mass and two sizes of the compact space,
the radius  $R$  of $S_1$ and the volume $(2\pi)^4 V$ of $T^4$.  
The three charges correspond to the number $Q_5$ of D-fivebranes
wrapped on $T^5$, the number $Q_1$ of D-strings wrapped on $S_1$
and the momentum $N/R$ flowing along $S_1$.
We use the notations of \hms  . 
We will be considering the limit
$ V , \gamma  $ large and $r_0$ small with $ V r_0^2 $ fixed
and $ r_0^2 e^{ 2 \gamma } $ fixed.
In this limit the black hole is more properly interpreted as
 a black fivebrane
in ten dimensions if the radius $R$ is also large or as 
a black four-brane in nine extended dimensions if 
the radius $R$ is small.
Alternatively we could think that it is a near extremal black
hole in five dimensions with only one charge, the other two
charges would be zero or small, of the order of the deviation
from extremality.
U-duality \townsend\ maps this solution  into
the solutions  describing the  near extremal limit of
black holes coming from  D-branes wrapped on 
$T^5$, or solitonic fivebranes wrapped on $T^5$ or 
fundamental strings winding along $T^5$, in summary, the
near extremal limit in five extended dimensions 
of all BPS black holes   that preserve 1/2
of the supersymmetries.

In this limit 
the mass of the black hole becomes \hms 
\eqn\massbh{ M_{BH} = {  R V Q_5 \over g \alpha'^3 } +   
{   R V r_0^2  \over 2  g^2 \alpha'^4 }
(\cosh 2 \alpha +\cosh 2 \sigma  )
=M_{bare} + M~,
}
where the first term is interpreted as the
the mass of the bare extremal  D-fivebranes.
The entropy reads, in this limit,
\eqn\entropybh{
S = {A_{10}\over 4 G_{10}} = {A_5\over 4 G_5} =
{ 2 \pi  R V  r_0^2 \over  g^{3/2} \alpha'^{7/2} } 
\sqrt{Q_5} \cosh \alpha  \cosh \sigma ~,
}
and the winding (of 1D branes)  and momentum charges are \hms
\eqn\charges{\eqalign{
   Q_1 & = { V r_0^2  \over 2 g \alpha'^3 } \sinh 2 \a , \cr
  n &= {  R^2V r_0^2 \over 2  g^2 \alpha'^4 } \sinh 2 \sigma ~.
}}
 It was shown in \hms\ that the
the classical black hole Hawking Beckenstein entropy
has the simple form
\eqn\smira{S= 2 \pi( \sqrt{ N_1} + \sqrt{  N_{ \bar 1}})
( \sqrt{ N_5} + \sqrt{  N_{ \bar 5}})( \sqrt{ n_L} + \sqrt{ n_R})~}
in terms of a collection of ($N_1,~N_{\bar 1},~N_5,~N_{\bar 5},~n_R,~n_L$)
non-interacting branes, anti-branes and strings. For notations
see ref. \hms . The limit we are considering corresponds
to $N_{\bar 5} =0$, while the limits previously considered involved
two of the anti-charges very  small \refs{\ghas,\hms } or all anti-charges
small \cama . 

In the case that the two charges \charges\ are small ($\alpha \sim \sigma \sim
0$) the Hawking temperature of the near extremal solution is constant with
the value
\eqn\thaw{
T_H^{-1} = 2 \pi \sqrt{ g \alpha' Q_5 }~.
}
When the 
deviation  from extremality becomes very small the classical 
thermodynamic description would break down and we expect that the 
physical temperature would drop to zero. One would estimate that
this happens when the mass above extremality becomes of the
order of the Hawking temperature $M \sim T_H $ \thaw .

General rotating black holes in five dimensions 
 were constructed in 
ref. \cveticnew . In the limit we are considering,
their entropy formula becomes
\eqn\entrorotbh{\eqalign{
S = &2 \pi  \sqrt{
Q_5 { r_0^4 V^2 R^2\over 4 g^3 \alpha'^7 }
 \left( \cosh \alpha \cosh \sigma +
\sinh  \alpha \sinh \sigma \right)^2
- {(J_1 + J_2 )^2 \over 4 } }~~ + \cr
+ & 2 \pi \sqrt{
Q_5 { r_0^4 V^2 R^2 \over 4 g^3  \alpha'^7 }
 \left( \cosh \alpha \cosh \sigma -
\sinh  \alpha \sinh \sigma \right)^2
- {(J_1 - J_2 )^2 \over 4 } }~.
}}

\newsec{Effective strings }

Before we say anything about D-branes let us calculate
the energy and entropy that a gas of strings
living on the fivebrane would have.
We consider  superstrings with 
 tension 
$T_{eff} = { 1 \over 2 \pi \alpha'_{eff} } $ and 
an effective number of degrees of freedom $c_{eff}$,
where $c_{eff} = N_B + { 1\over 2 } N_f $ is the effective
central charge in the static gauge, which we assume to be the
same for left and right movers, since the central charge is
not 12 there are potential problems for constructing a superstring
theory.
For the moment
we will work in this gauge and we will not worry about
the problems with Lorentz covariance that might arise.
All we  will do now is to  rewrite the
entropy formula in a suggestive way, inspired in this
effective string idea. Without much justification
we will take this string to be non interacting.
We take this gas to have total mass $M$, and
some net momentum and winding ($n$,$m$)  along the $S^1$
direction.
The entropy of this gas comes mainly from a configuration
in which we have just a single long string.
The right and left moving momenta
of
this long string
in the $S^1$ direction are
\eqn\plr{
p_{R,L} = { n \over R } \pm { m R \over  \alpha'_{eff} } ~.
}
The Virasoro constraints are
\eqn\mass{
M^2 = p_R^2 + {4 \over  \alpha'_{eff} } N_R~,~~~~~~~~~~~~~~~~
M^2 = p_L^2 + {4 \over  \alpha'_{eff}  } N_L~.
}
where $N_{L},N_R$ are the left and right total oscillator levels.
Using the number of physical degrees of freedom that we mentioned
above, the entropy becomes
\eqn\entrofi{
S = 2 \pi \sqrt{ c_{eff}/6 } \left(
\sqrt{ N_L}  +
\sqrt{ N_R} \right)~.
}
Let us  identify the mass of this string gas with the
the second term in the mass of the black hole \massbh ,
the momentum $n$ with the charge $n$  in \charges\ and the
the winding number $m$ with the charge $Q_1$ as
\eqn\qone{ Q_1  = m { g \alpha' \over
\alpha'_{eff} } ~}
which follows from comparing the black hole
 BPS mass formula  and  \mass\ for
strings with pure winding.
Now, using \mass \plr\ we can calculate $N_{R,L}$
\eqn\numbers{
N_{R,L} = { r_0^4 V^2 R^2 \over 4 g^3 \alpha'^7  }
{  \alpha'_{eff} \over  g \alpha' }
 \left( \cosh \alpha \cosh \sigma \pm
\sinh  \alpha \sinh \sigma \right)^2
}
so that \entrofi\ would reproduce \entropybh\ if
\eqn\ceff{
{ c_{eff} \over 6 }.{ \alpha'_{eff} \over  g \alpha'  } = Q_5 ~.
}
The Hagedorn temperature for this string is $T^{-1}_H = 2 \pi 
\sqrt{ \alpha'_{eff} c_{eff}/6 } $ which, using \ceff\
becomes \thaw .

Note that all we have done in this section is to make a
change of variables and rewrite the entropy formula \entropybh\
in the suggestive form \entrofi . The entropy presented in
this way strongly suggests a microscopic description in terms of strings
living on the fivebrane. This gives a very natural explanation for
the constant temperature \thaw . 

\newsec{ D-brane picture}

We will now show that strings obeying \qone \ceff\
arise naturally in the D-brane description of black holes.
In the D-brane picture,
extremal black holes can be viewed as a collection of $Q_5$ 
 D-fivebranes wrapping on
$T^4\times S^1$, $Q_1$ D-strings wound along $S^1$ with some
momentum $N$ flowing along the direction of the strings
\refs{\ascv,\cama}.
In \dgl\ the D-strings were pictured as instantons in
the $U(Q_5)$ gauge theory of the fivebrane, more precisely,
the D-strings are the zero size limit of these  instantons.
In \vins\ it was argued  that the moduli space of one of these
instantons corresponds to an $N=4$ superconformal field theory
with $ 4 Q_5$ bosonic and $ 4 Q_5 $ fermionic degrees of freedom.
So we can say that they are ``instanton strings''  with
$ 4 Q_5$ bosonic and $ 4 Q_5 $ fermionic physical oscillators.
The same conclusion can be achieved with the analysis in
 \cama\ where this number of degrees of freedom
is related to open strings going between the D-string and the
$Q_5$
D-fivebranes. We are now concentrating in
just one D-string at a time and  we are  considering only
the massless  modes of the D-string.
The inverse  tension of this string  is just that of
the D-string $ \alpha'_{eff} =  g \alpha'  $, and the 
effective central charge is $c_{eff} = 6 Q_5 $ so that
\qone \ceff\ are  satisfied. 
One puzzling aspect of this picture is that the 
string tension is quite large for weak coupling so that
one might wonder why just a gas of fundamental strings is not giving 
a bigger contribution. One answer to this is to
note that for large $Q_5$, and energies such that $M_{bare} \gg
M \gg {1 \over \sqrt{g \alpha' } }$,  these ``instanton strings''
 have much more
entropy. 
This suggests that the entropy formula 
\entrofi\ will be valid only for energies bigger than
 $M \sim {1 \over \sqrt{g \alpha' } }$.
However, from the classical black hole theory we would 
expect it to break down only when $M \sim T_H $ \thaw . 
These two scales are very different for large $Q_5$. 
In \sm\ similar problems were solved by proposing that, due
to the way the branes are wrapped, the effective quantum 
of momentum for excitations moving along the compact direction
 is changed $ 1/R \rightarrow 1/R Q_5 $. In a similar fashion
we expect 
that a more correct picture should be that these 
``instanton strings'' break into $Q_5$ fractional instantons
with instanton number $1/Q_5$. Indeed it 
 was found in \thooft\ that there are
 solutions of U($Q_5$) gauge theories on a torus 
with fractional instanton number $ 1/Q_5$.
These fractional strings would have 
inverse string
tension $\alpha'_{eff} = Q_5 g \alpha' $ and $c_{eff} =6$ so
that \qone \ceff\ continues to be satisfied\foot{
It would be very interesting to find the relation between these
effective strings and the ones in \verlinde \schwarz .}
A picture of strings with renormalized tension
was advocated in \lawi\ to account for the entropy in the BPS case.
These strings loosely resemble those proposed by Susskind \lsrutgers .
Of course there is another problem which is that of strong coupling.
It  is apparently present in all discussions of non-extremal
black holes and makes these formulas look even more puzzling.

As a further check on the effective string
 picture we will calculate
the entropy for the case with angular momentum
in the extended 1+4 dimensional
space.
The worldsheet theory, in the static gauge, 
of these ``instanton strings''  is 
 an N=(4,4) superconformal field theory.  
It was argued in \refs{\spn,\vbd } that
the angular momentum is carried by the fermions of
the N=(4,4) superconformal field theory. More precisely,
the N=(4,4) algebra  has a $SU(2)_R \times
SU(2)_L $ R-symmetry which, as argued in \spn , is linked
 to the decomposition of the spatial rotations as
 $ SO(4) = SU(2)_R \times SU(2)_L $.
In SO(4) language the angular momentum is  characterized by
two eigenvalues $J_1, J_2 $ corresponding to the angular
 momenta on  two orthogonal planes and, in terms of
the U(1) charges  ($F_R, F_L$)  of the $N=4$ algebra,
 they read \spn
\eqn\ang{
J_1 = { 1\over 2} ( F_R + F_L )~,~~~~~~~~~~~~~~~~~
J_2 = { 1\over 2} ( F_R - F_L )~.
}
Using the same arguments as \refs{\spn,\vbd }
we find that the
net effect will be
to replace in \entrofi
\eqn\tilns{
N_L \rightarrow \tilde N_L =
N_L - {(J_1 - J_2 )^2 \over 2 c_{eff}/3 } ~,~~~~~~~~~~~~~~~~
N_R \rightarrow \tilde N_R = N_R - {(J_1 + J_2 )^2 \over 2 c_{eff}/3  }
}
so that
\eqn\srot{\eqalign{
S =& 2 \pi \sqrt{c_{eff} \over 6} \left(
\sqrt{ \tilde N_R}
+
\sqrt{ \tilde N_R} \right) \cr
=& 2 \pi
 \sqrt{ { c_{eff}\over 6}  
N_R - {(J_1 + J_2 )^2 \over 4 } } +  2 \pi
 \sqrt{  { c_{eff}\over 6} 
N_L - {(J_1 - J_2 )^2 \over 4 }}~,
}}
which, using \numbers ,\ceff\ reduces to \entrorotbh .

\newsec{Discussion}

In summary, a remarkable formula \entrofi\ has been found for the
entropy on near extremal fivebranes
  using a D-brane inspired  picture in terms of strings living
on the fivebrane.
The  near 
extremal limit considered here is more generic than the ones
considered in
\refs{ \ghas,\hms,\vbd,\cama }, unfortunately the 
objects we are dealing with are not so well defined.
The energy dependence of the near extremal entropy is
well reproduced by a gas of strings. The precise
numerical coefficient requires a more specific property
\ceff\ of   this effective  string theory.
 This  property 
is precisely obeyed by ``instanton strings''  arising
in the D-fivebrane gauge theory. In the PBS limit these
strings were used to explain the entropy \ascv .
It will be interesting to study more precisely these strings
in the spirit of \refs{\schwarz,\verlinde}, we see that some more
information can be gained by studying near extremal black holes.
Under U-duality the D-fivebrane becomes a solitonic fivebrane
and the D-strings fundamental strings. It will be interesting 
to connect the present picture with the dilaton gravity studies
\dilgra .

It is very interesting that the near extremal black hole
entropy
can be viewed as  the entropy as a gas of particles
 when two of the charges
are large \refs{\ghas}  and as the entropy 
of a  gas of strings  when only one of the charges is large.
The general case, which includes the Schwarschild black hole,
 might involve also strings or some other
extended object.

It seems straightforward to generalize these arguments to 
four dimensions using the results in \refs{\jmas,\hlm} to 
the case where two of the anti-charges are different from zero.
These black holes are U-dual to the near extremal limit
of Sen's electrically charged black holes \sen . 
Indeed the thermodynamics features of \sen\ suggest an effective 
string picture 
 since the
near extremal Hawking temperature of Sen's black holes 
is constant (therefore the entropy is linear in the mass difference
$M_{BH}- M_{extr}$).

A word of caution should be said: we have overlooked two 
problems, one is how to get a fully consistent
free superstring theory and the other is how they 
manage to be weakly coupled. Hopefully the resolution to these
problems will lead to a  more precise picture.
It is however very interesting to have found a black hole where
the entropy is indeed given by a gas of strings.

{\bf Acknowledgments}

I would like to thank specially C. Callan for many comments, and
also 
G. Horowitz, I. Klebanov and  A. Strominger
for very interesting discussions.
I also thank V. Balasubramanian, F. Larsen, 
S. Ramgoolam and E. Sharpe.
This work was supported in part by DOE grant DE-FG02-91ER40671.

\listrefs

\bye